\documentclass[%
 reprint,
 amsmath,amssymb,
 aps,
superscriptaddress]{revtex4-2}
\usepackage[colorlinks=true, linkcolor = blue, hyperfootnotes=false, citecolor = blue]{hyperref}
\usepackage{graphicx}
\usepackage{dcolumn}
\usepackage{bm}
\usepackage[dvipsnames]{xcolor}
\usepackage{gensymb}

\begin{document}

\preprint{APS/123-QED}

\title{A controllable anti-P-pseudo-Hermitian mechanical system and its application}

\author{Yanzheng Wang}
\thanks{These authors contributed equally to this work.}
\affiliation{Huanjiang Laboratory, Zhejiang University, Zhuji 311800, PR China}
\affiliation{Department of Mechanical and Aerospace Engineering, University of Missouri, Columbia, MO 65211, USA}
\author{Jianlei Zhao}
\thanks{These authors contributed equally to this work.}
\affiliation{Department of Mechanics and Engineering Science, Peking University, Beijing 100871, PR China}
\author{Qian Wu}
\affiliation{Department of Mechanical and Aerospace Engineering, University of Missouri, Columbia, MO 65211, USA}
\author{Xiaoming Zhou}
\affiliation{Key Laboratory of Dynamics and Control of Flight Vehicle of Ministry of Education, School of Aerospace Engineering, Beijing Institute of Technology, Beijing, 100081, China}
\author{Heng Jiang}
\affiliation{Key Laboratory of Microgravity, Institue of Mechanics, Chinese Academy of Sciences, Beijing 100190, China}
\author{Weiqiu Chen}
\affiliation{Department of Engineering Mechanics, Zhejiang University, Hangzhou 310007, PR China}
\author{Mu Wang}
\affiliation{National Laboratory of Solid State Microstructures, School of Physics, and Collaborative Innovation Center of Advanced Microstructures, Nanjing University, Nanjing 210093, PR China}
\author{Guoliang Huang}
\email[Corresponding author:]{guohuang@pku.edu.cn}
\affiliation{Department of Mechanics and Engineering Science, Peking University, Beijing 100871, PR China}

\date{\today}

\begin{abstract}

A novel anti-P-pseudo-Hermitian mechanical system that integrates piezoelectric actuators and sensors with non-reciprocal coupling into mechanical beams is proposed. This configuration enables the system to exhibit programmable exceptional points (EPs), which are critical for enhancing sensitivity in sensing applications. Our theoretical analysis, supported by numerical simulations and experimental validation, demonstrates the system's capability to detect minute mass variations and identify surface cracks with high precision. This advancement not only contributes to the field of non-Hermitian physics but also paves the way for the development of next-generation mechanical sensors leveraging EP physics.

\end{abstract}

\maketitle

In the past two decades, there has been a growing interest in investigating non-conservative systems, particularly in connection with quantum mechanics, after the realization that some non-Hermitian Hamiltonians exhibit entirely real spectra as long as the system preserves the pseudo-Hermitian symmetry \citep{makris2008beam,mandal2021symmetry,heiss2012physics,wu2024understanding,bender1998real}, which includes PT-symmetric, anti-PT-symmetric, and P-pseudo-Hermitian systems. Naturally, anti-P-pseudo-Hermitian system shares similar properties. In a PT-symmetric system, the non-Hermitian Hamiltonian $\mathbb{H}$ remains invariant under combined parity (P) and time-reversal (T) operations, satisfying $\mathbb{H}=\text{PT}\mathbb{H}(\text{PT})^{-1}$ \citep{bender2002complex}.  In contrast, an anti-PT-symmetric system obeys $\mathbb{H}=-\text{PT}\mathbb{H}(\text{PT})^{-1}$, where the Hamiltonian changes sign under the same operations. For P-pseudo-Hermitian system, the Hamiltonian $\mathbb{H}$ under the parity operation returns to its conjugate transpose \citep{mostafazadeh2003pseudo}, i.e. $\mathbb{H}^\dagger=
\text{P}\mathbb{H}\text{P}^{-1}$, where “$\dagger$” denotes the Hermitian conjugate transpose. The anti-P-pseudo-Hermitian counterpart satisfies $\mathbb{H}^\dagger=-\text{P}\mathbb{H}\text{P}^{-1}$, requiring the conjugate transpose of the Hamiltonian to flip sign under parity. However, the realization of an anti-P-pseudo-Hermitian system typically requires non-reciprocal and purely imaginary couplings between eigenmodes \citep{ashida2020non}. Due to this fundamental challenge, the design of the anti-P-pseudo-Hermitian symmetry has never been attempted in any systems so far. 

The well-known PT symmetry \citep{ozdemir2019parity} and anti-PT symmetry \citep{yang2017anti} have received great attention across different physical platforms for various applications \citep{kononchuk2022exceptional,ren2017ultrasensitive,chen2017exceptional,de2021exceptional}. Theoretically, PT symmetry requires incorporating energy gain and its delicate balance with the loss \citep{ghosh2021classical, meng2024exceptional,ruter2010observation}. In optical and electric systems, fascinating behaviors such as unidirectional invisibility \citep{lin2011unidirectional} and enhanced sensitivity \citep{de2022non,hodaei2017enhanced} have been experimentally revealed at the EPs in PT-symmetric systems. On the other hand, most mechanical systems achieve the gain effect from artificial forces caused by optical or electronic sources \citep{rosa2021exceptional,wu2019asymmetric,thomes2024experimental,xu2015mechanical}, and the application of small mass or defect sensing is conceptually explored \citep{jin2024exceptional,cai2022exceptional,yuan2022exceptional}. The anti-PT-symmetric system does not require the simultaneous presence of gain and loss but poses a more challenging constraint on the purely imaginary coupling of eigenmodes \citep{ fan2020antiparity,de2019high,li2019anti}. This makes anti-PT symmetry difficult to engineer in practice, and only realized in optical, electrical, and thermal diffusive systems through the direct \citep{choi2018observation,zhang2020synthetic,li2019anti} or indirect \citep{wei2021anti,zhang2019dynamically} coupling strategies. However, there have been no reports on the study of anti-PT-symmetric mechanical systems, not to mention the anti-P-pseudo-Hermitian symmetry, which further requires non-reciprocal coupling. Despite the exotic physical properties conferred by the anti-P-pseudo-Hermitian systems \citep{zheng2020efficient}, the physical construction of the non-reciprocal coupling remains elusive, the microstructure design of which is a major challenge in this field. 

\begin{figure}
\includegraphics[width=\linewidth]{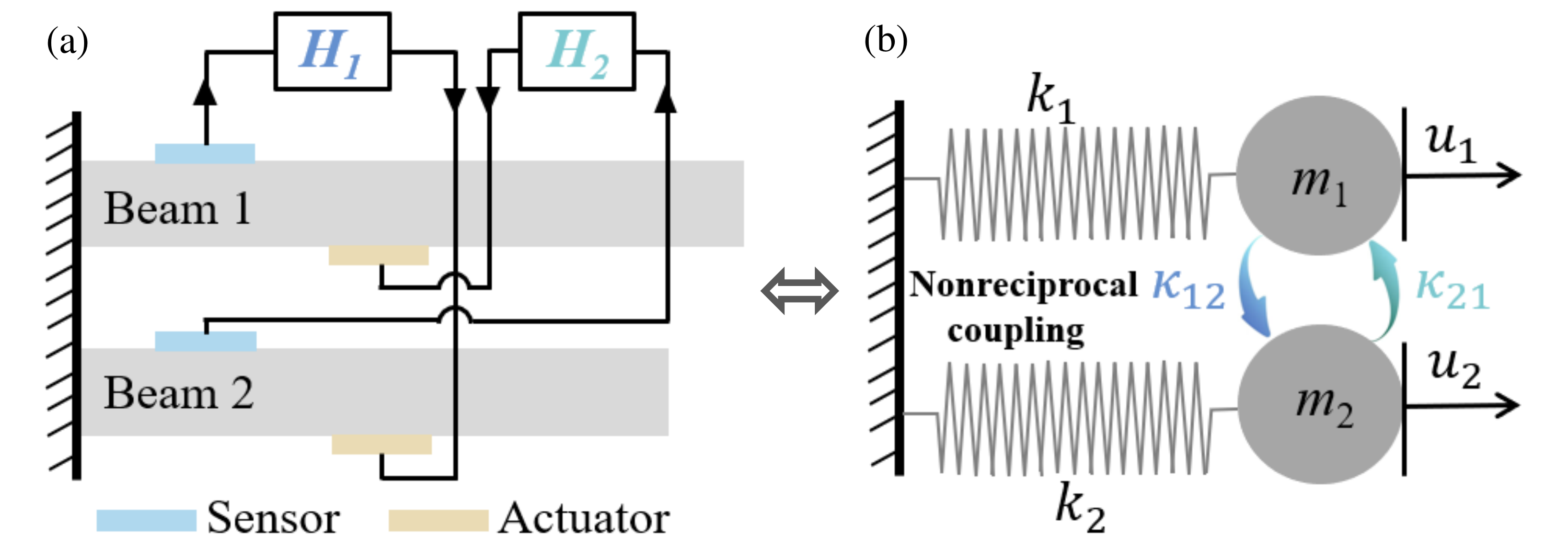}
\caption{Design of a controllable anti-P-pseudo-Hermitian system. (a) The physical illustration of the system consisting of two cantilever beams connected through piezoelectric sensors and actuators controlled by two transfer functions, $H_1$ and $H_2$, to realize non-reciprocal coupling, (b) the equivalent mass-spring model with nonreciprocal coupling.}
\label{fig1}
\end{figure}

In this letter, we introduce and prove a universal anti-P-pseudo-Hermitian mechanical system consisting of two simple cantilever beams with non-reciprocal couplings. To construct the non-reciprocal coupling, we propose an innovative approach to electrically connect the two beams through piezoelectric patches controlled by two different transfer functions,  as illustrated in Fig. \ref{fig1}(a). Thanks to the programmability of the transfer functions, the non-Hermitian system can be easily reduced into either PT symmetry or anti-PT symmetry counterparts by tuning the non-reciprocal coupling into purely real or imaginary reciprocal coupling through microcontrollers \citep{chen2021realization, wang2024non, wu2023engineering, wu2023active}. It should be emphasized that our proposed structural design fundamentally differs from existing approaches that achieve non-Hermitian PT-symmetric systems through reciprocal coupling between electrical and mechanical modes \citep{rosa2021exceptional,wu2019asymmetric,thomes2024experimental}. For the present design, the two mechanical modes are coupled in a controllable manner via a sensor-actuator-feedback control loop, where the piezoelectric patches serve as sensors and actuators, respectively. For the anti-P-pseudo-Hermitian system, the two mechanical beams and transfer functions do not need to be identical. We theoretically prove that the anti-P-pseudo-Hermitian system is capable of exhibiting programmable EPs and validate this theoretical framework numerically and experimentally. This controllable feature ensures that the system consistently operates around the EP, even when beam properties vary due to degradation in working conditions. To demonstrate the ultrasensitive sensing application of the system, we study the robust ability to detect minor changes in mass and identify surface cracks. Our work lays a solid foundation for developing next-generation mechanical sensing technologies.

To analyze the dynamic behavior of the non-Hermitian system, the lengths of the two beams are denoted as $L_1$ and $L_2$, respectively. In the system, the transfer functions are defined as $H_1=V_{a1}/V_{s2}$ and $H_2=V_{a2}/V_{s1}$, respectively. The piezoelectric actuators will elongate or contract in response to the applied voltage, thereby adjusting the effective bending stiffnesses of the corresponding beams. The piezoelectric patches, functioning as sensors, detect the bending deformation of the bonded beams by generating voltages $V_{s1}$ and $V_{s2}$, which are proportional to the elongation or contraction of the beams' top surfaces. These sensing voltages are then amplified through the electric circuits and applied to the piezoelectric actuators ($V_{a1}$ and $V_{a2}$) on the opposite beams, thereby inducing controlled bending responses. The elastodynamic equations in the frequency domain are expressed as

\begin{equation}\label{eq1}
\omega^2\mathbf{\Psi}^{mq}=\mathbb{H}^{mq}_\text{con}\mathbf{\Psi}^{mq}
\end{equation} 
where $\omega$ denotes the system's frequency, and $\mathbf{\Psi}^{mq} = [A_{1}^m, A_{2}^q]$ represents the vibration mode amplitudes of the $m^{th}$ vibration mode of beam 1 and the $q^{th}$ vibration mode of beam 2. The corresponding dynamic matrix can be derived as
\begin{equation}\label{H_1}
\mathbb{H}^{mq}_\text{con}=
\begin{bmatrix}
\cfrac{D_\text{eff1}^m}{m_\text{eff1}^m} &-H_1 P_1\\
-H_2 P_2&\cfrac{D_\text{eff2}^q}{m_\text{eff2}^q}\\
\end{bmatrix}
\end{equation} 
where $D_\text{eff1}^{m}$, $m_\text{eff1}^{m}$, $D_\text{eff2}^{q}$, and $m_\text{eff2}^{q}$ represent the effective bending stiffness and the effective mass for the $m^{th}$ vibration mode of beam 1 and the $q^{th}$ vibration mode of beam 2, respectively. The effective bending stiffness can be complex due to physical and active damping effects. $P_1$ and $P_2$ represent the coupling coefficients determined by the beams' geometries and eigenmodes, which are real-valued and have the same sign if the eigenmodes of the two beams have similar mode profiles. A detailed derivation of Eqs. (\ref{eq1}) and (\ref{H_1}) are provided in Supplementary Materials II. The controllable coupling between the two beams is fully determined by the transfer functions ($H_1$ and $H_2$), which can be selected in any form, such as the reciprocal or nonreciprocal coupling in terms of the real or imaginary numbers. Therefore, the suggested system is electronically reconfigurable to investigate various non-Hermitian structures, encompassing PT-symmetric, anti-PT-symmetric, and anti-P-pseudo-Hermitian systems. A detailed analysis of the symmetry of the Hamiltonian is provided in Supplementary Materials III. 

To achieve the anti-P-pseudo-Hermitian mechanical system, the coupling coefficients must be (i) purely imaginary and (ii) not equal. Thus, the transfer functions have to meet $H_1=i \overline{H}_1$ and $H_2=i \overline{H}_2$, where $\overline{H}_1$ and $ \overline{H}_2$ are real values. Performing the unitary transformation on the dynamic matrix as $\mathbb{\overline{H}}^{mq}_\text{con}=\mathbf{U}^\dagger\mathbb{H}^{mq}_\text{con}\mathbf{U}$ ensures that the characteristic polynomial and the corresponding eigenvalues of the dynamic matrix remain unchanged \citep{strang2022introduction}. Thus, we can still refer to the new system with the transformed dynamic matrix as the anti-P-pseudo-Hermitian system. Here, the unitary matrix is chosen as $\mathbf{U}$=diag(1,$i$). For the new system, the transfer functions become real and can be represented as $H_1=-\overline{H}_1$ and $H_2=\overline{H}_2$. Thus, we can conclude that the non-reciprocal coupling is the only necessary condition for realizing the anti-P-pseudo-Hermitian system.

To quantitatively describe the dynamics of the continuum-based system, the dynamic matrix for two coupled discrete mechanical oscillators, illustrated in Fig. \ref{fig1}(b), is expressed as
\begin{equation}\label{H_2}
\mathbb{H}_\text{cmt}=
\begin{bmatrix}
-\omega_1^2-i\gamma_1
&\kappa_{12}\\
\kappa_{21}&-\omega_2^2-i\gamma_2\\
\end{bmatrix}
\end{equation} 
where $\omega_1^2 = k_1/m_1$ and $\omega_2^2 = k_2/m_2$. By equating the dynamic matrix of the two systems, we have
\begin{equation}\label{effective}
\begin{aligned}
k_1=\text{Re}(D_\text{eff1}^m),m_1=m_\text{eff1}^m, \\ k_2=\text{Re}(D_\text{eff2}^q),m_2=m_\text{eff2}^q, \\
\gamma_1=\cfrac{\text{Im}(D_\text{eff1}^m)}{m_\text{eff1}^m}, \gamma_2=\cfrac{\text{Im}(D_\text{eff2}^q)}{m_\text{eff2}^q},
\end{aligned}
\end{equation} 
and the non-reciprocal coupling strengths are
\begin{equation}
\kappa_{12}=i\overline{H}_1P_1,\kappa_{21}=i\overline{H}_2P_2,
\end{equation}
or, alternatively,
\begin{equation}
\kappa_{12}=-\overline{H}_1P_1,\kappa_{21}=\overline{H}_2P_2.
\end{equation}

\begin{figure}
\centering
\includegraphics[width=\linewidth]{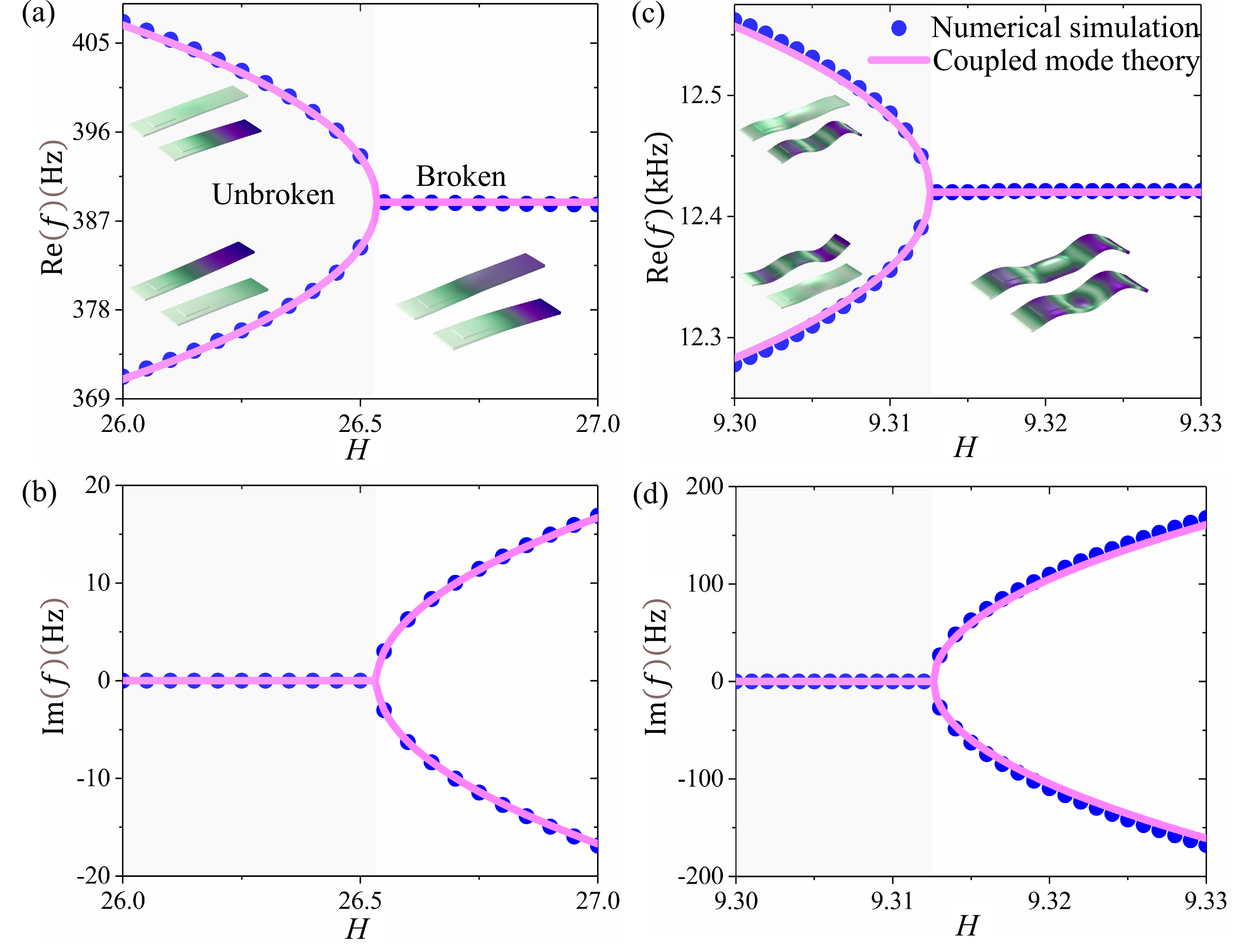}
\caption{The EPs for the vibration of the anti-P-pseudo-Hermitian mechanical system. The real parts (a,c) and the imaginary parts (b,d) of the eigenfrequencies of the system with the transfer function. (a,b) and (c,d) show the results for the first- and fourth-order flexual vibration modes, respectively.}
\label{fig2}
\end{figure}

 Following the design guideline, we physically construct an anti-P-pseudo-Hermitian mechanical system using two electrically coupled cantilever beams made of aluminum with different sizes: $50 \times 10 \times 0.8$ mm$^3$ and $40 \times 10 \times 0.8$ mm$^3$. The system includes PZT-5H patches acting as sensors and actuators with dimensions of $10 \times 6 \times 0.52$ mm$^3$. The sensors are placed at $L_i/8$ and the actuators are located at $L_i/2$ along the length of each beam $i=1,2$. The discussion on the selection of the positions of the sensors and actuators is provided in Supplementary Materials IV. For simplicity, we consider a case where $\overline{H}_1=\overline{H}_2=H$ and beam damping is negligible in this study. To understand the working mechanism, the system’s eigenfrequencies near the EP can be analytically derived based on the coupled mode theory as
\begin{equation}\label{fH}
f=A\pm B\sqrt{H_0-H}
\end{equation} 
where $A=\sqrt{(f_1^2+f_2^2)/2}$ and $B=\sqrt{(f_1^2-f_2^2)/(f_1^2+f_2^2)/2}/H_0$, $f_1$ and $f_2$ are the natural frequencies of the beam 1 and the beam 2 without coupling and $H_0$ is the transfer function to reach the EP. Here, the condition $\lvert(H_0 - H)/H_0\rvert \ll 1$ is adopted, and the values for $H_0$, $A$, and $B$ are numerically determined. The details are provided in the Supplementary Materials IV.

Figure \ref{fig2} shows the evolution of the predicted real (Fig. \ref{fig2}(a)) and imaginary parts (Fig. \ref{fig2}(b)) of eigenfrequencies of the first-order vibration of the anti-P-pseudo-Hermitian system with the increase of $H$. The numerical results on the actual system agree well with those from the coupled mode theory described by Eq. (\ref{fH}). It can be observed that the eigenfrequency splitting around the EP follows a square-root dependence on the change of $H$. As $H$ increases, the eigenfrequencies of the shorter beam decrease due to mechanical softening caused by the coupling. Conversely, the eigenfrequencies of the longer beam increase because of mechanical stiffening. The real parts of the two eigenfrequencies gradually approach each other and eventually merge to form an EP when the coupling reaches a critical value. Before the EP (illustrated in the gray area), the system remains in the unbroken phase with real-valued eigenfrequencies, indicating a balance between the positive and negative work done by the actuators on the two beams. Beyond this point (illustrated in the white area), the system enters the broken phase, and the imaginary parts start to bifurcate into two branches, where the eigenfrequencies become complex conjugates and the system becomes unstable due to an imbalance in work done. The vibration eigenmodes before and after the EP are also inserted in Fig. \ref{fig2} further to reflect this vibrational energy exchange in the non-Hermitian system. Figures \ref{fig2}(c)-\ref{fig2}(d) present the evolution of the fourth-order eigenfrequencies with the change in $H$, similar to those discussed in the first-order eigenfrequencies. Notably, achieving the EP of the fourth-order eigenfrequencies requires a smaller value of $H$, making it easier to implement experimentally.

\begin{figure}
\includegraphics[width=\linewidth]{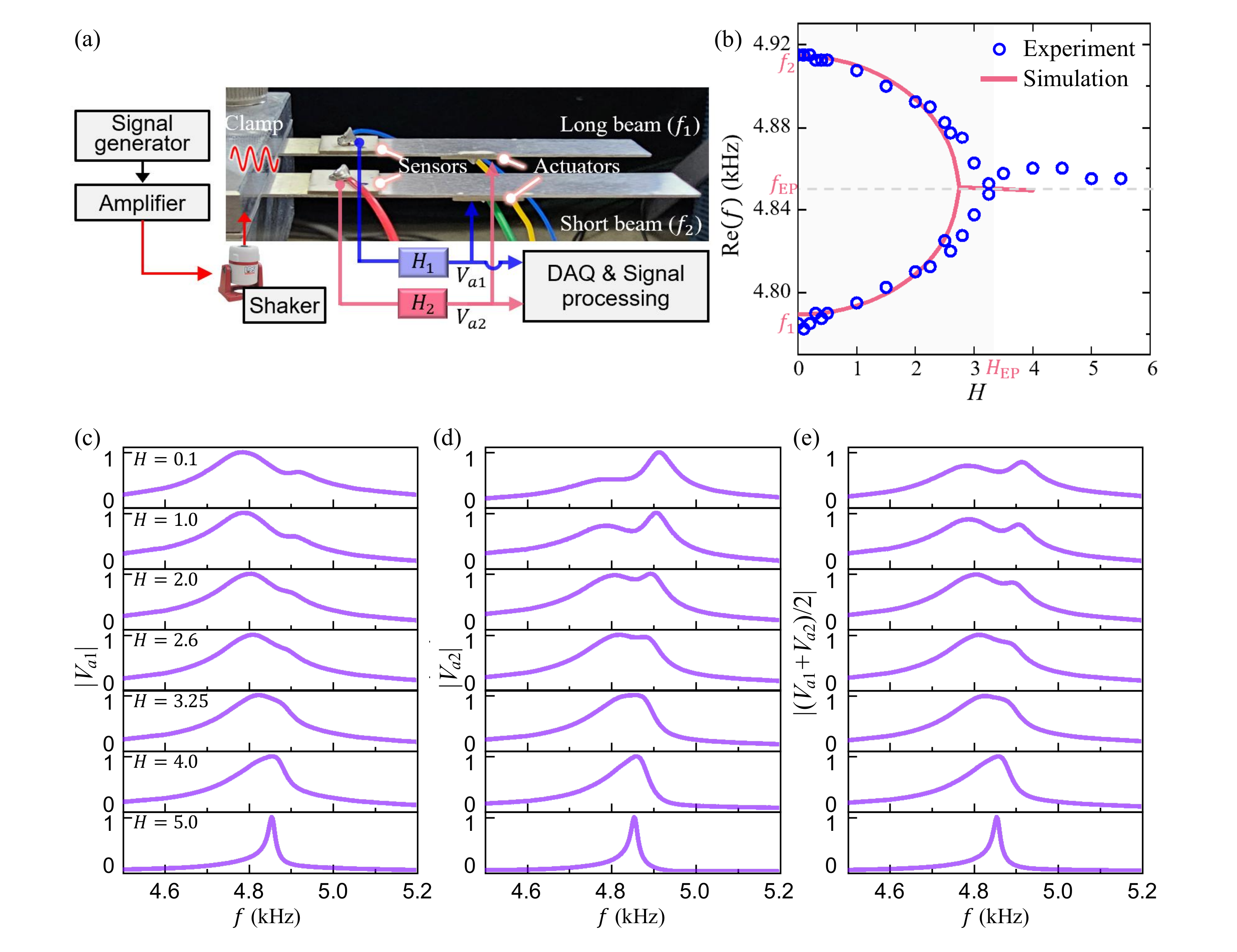}
\caption{Experimental demonstration of the EPs in the anti-P-pseudo-Hermitian mechanical system. (a) Schematic of experiment setup. (b) Experimental proof of the EPs for the fourth-order bending mode. (c-e) Frequency spectrum of $V_{a1}$, $V_{a2}$ and their arithmetic mean $(V_{a1}+V_{a2})/2$ for several representative values of $H$ near the fourth-order eigenfrequency. Each output voltage was independently normalized by its maximum value.}
\label{Experiment}
\end{figure}

\begin{figure}
\includegraphics[width=\linewidth]{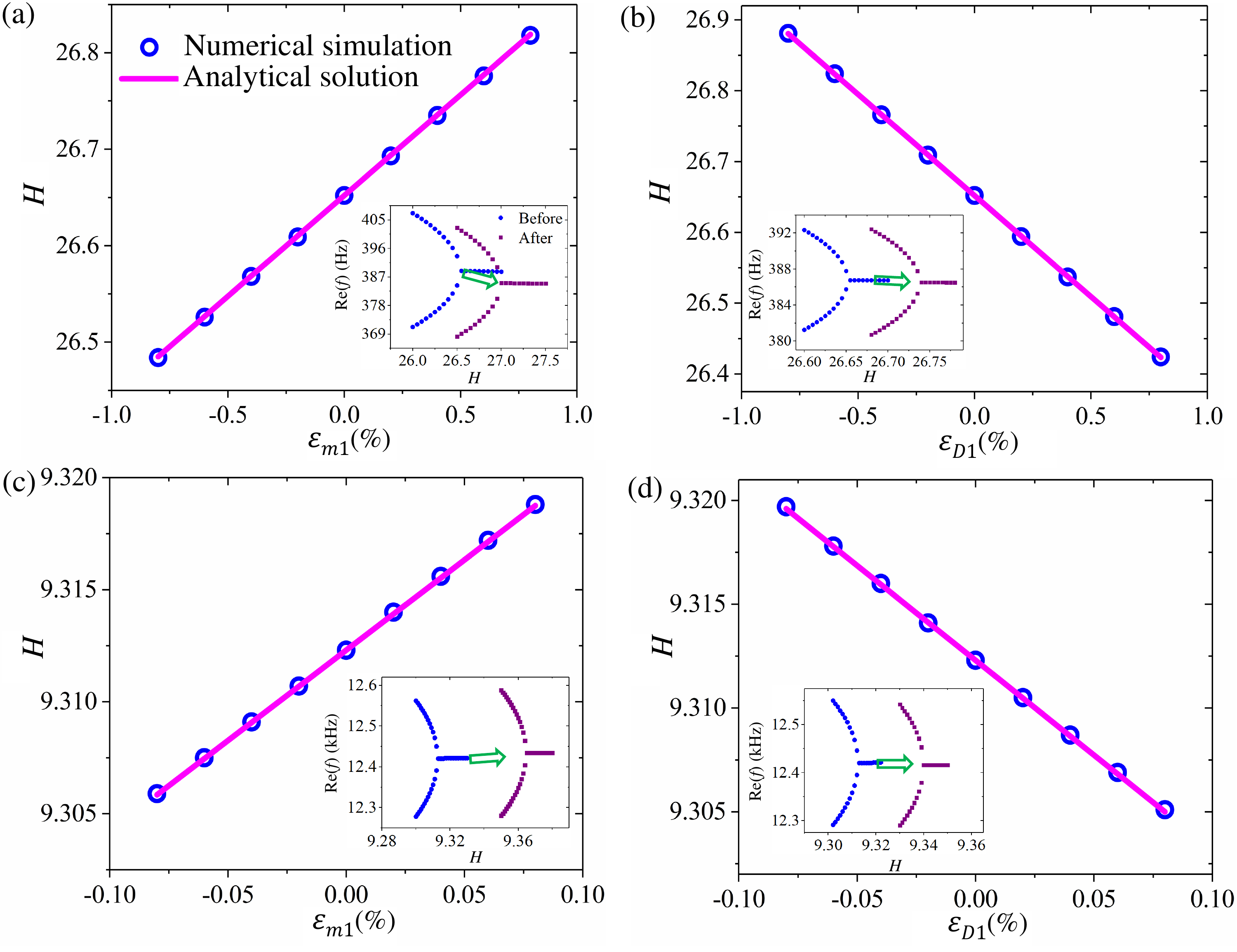}
\caption{Controllability of EPs for the non-Hermitian mechanical system. The programming of the transfer function ensures the exceptional point (EP) of the system following the mass perturbation (a, c) and the stiffness perturbation (b, d). (a,c) correspond to the first-order mode and (b,d) correspond to the fourth-order mode. The inserted figures show the system can be adapted to approach the EP by changing $H$ after adding a mass on beam 1 (a, c) or reducing the stiffness of beam 1 (b, d).}
\label{fig3}
\end{figure}

To validate the design of the anti-P-pseudo-Hermitian mechanical system, the experimental testing is conducted using integrated sensing, actuation, and closed-loop feedback control. As shown in Fig. \ref{Experiment}(a), the two mechanical beams are clamped onto a shaker, which is driven by a frequency-sweep excitation signal generated by a signal generator and amplified through a power amplifier (see Supplemental Material XII for experimental details). Figure \ref{Experiment}(b) shows the experimentally measured eigenfrequency evolution of two fourth-order bending modes (blue circles) as a function of the transfer function $H$, where $f_1$ and $f_2$ denote the eigenfrequencies of the long and short beams, respectively. As $H$ increases, these two eigenfrequencies gradually coalesce into a single degenerate eigenfrequency $f_{EP}$. For comparison, the numerical simulation is also performed. It is found that the experimentally observed critical value $H_{EP}$ is higher than the one obtained from the numerical prediction (solid line), primarily due to the reduced actuation efficiency of the piezoelectric patch caused by the presence of the relatively thick conductive adhesive bonding layer. Figures \ref{Experiment}(c)-\ref{Experiment}(e) show the frequency-response curves at several representative values of $H$, from which the eigenfrequencies are extracted. The outputs used include $V_{a1}$ and $V_{a2}$ from the two mechanical beams, respectively, and their arithmetic mean. It can be found that the mean of $V_{a1}$ and $V_{a2}$ provides the clearest indication of the evolution of the two eigenfrequencies to converge to the EP, thus confirming the validity of our anti-P-pseudo-Hermitian system design  (Fig. \ref{fig1}(a)).

Controllability is another distinct feature in the design of adaptive non-Hermitian systems enabling dynamic tuning of coupling strengths to sustain their performance near EPs, making them well-suited for practical sensing applications. For example, the proposed system can maintain maximum sensitivity by re-adjusting coupling in response to minor perturbations in material properties. Theoretically, the Hamiltonian obtained in Eqs. (\ref{H_1}) and (\ref{H_2}) for any mechanical perturbations can be interpreted as a perturbation of eigenfrequency 
\begin{equation}\label{H_e}
\mathbb{H}=
\begin{bmatrix}
-\omega_1^2+\epsilon_1
&\kappa_{12}\\
\kappa_{21}&-\omega_2^2+\epsilon_2\\
\end{bmatrix}
\end{equation} 
where $\epsilon_1$ and $\epsilon_2$ represent minor perturbations in beams 1 and 2, respectively. For example, in the case of the small perturbations applied solely to beam 1, the relationship between the transfer function and the micro-perturbations required to approach the EP can be derived as
\begin{equation}\label{Pb}
H=\beta_1 \epsilon_{m1}-\beta_1 \epsilon_{D1}+\beta_2-\beta_1
\end{equation} 
where $\epsilon_{m1}=(m_\text{eff1}-m_\text{eff1}^0)/m_\text{eff1}^0 \ll 1$ and $\epsilon_{D1}=(D_\text{eff1}-D_\text{eff1}^{0})/D_\text{eff1} \ll 1$ denote the mass and stiffness perturbations, respectively, where $m_\text{eff1}^0$ and $D_\text{eff1}^{0}$ are the initial effective mass and bending stiffness of beam 1, respectively. $\beta_i$ ($i=1,2$) are the parameters to be determined numerically. The detailed derivation is provided in Supplementary Materials V. From Eq. (\ref{Pb}), it can be deduced that the transfer function should be electronically adjusted linearly with small perturbations in mass and stiffness to re-approach the new EP. For the first-order mode, a small increase in the mass density of beam 1 necessitates a reduction in the transfer function to re-approach the EP, as shown in Fig. \ref{fig3}(a). Conversely, a minor increase in the stiffness of beam 1 requires the transfer function to increase, as illustrated in Fig. \ref{fig3}(b). For the fourth-order mode, the tunable behavior of the transfer function is similar, as depicted in Figs. \ref{fig3}(c) and \ref{fig3}(d). The predicted results are in excellent agreement with the numerical simulations. The figures inserted in Fig. \ref{fig3} further demonstrate how the system can be adapted to approach the EPs by adjusting the transfer function after adding mass to beam 1 or decreasing its bending stiffness. The size of the mass is selected as $2\times 2\times 2$ mm$^3$, whose initial density is assumed as $1000$ kg/m$^3$. The reduced stiffness for beam 1 is $0.3\%$.

\begin{figure}
\includegraphics[width=\linewidth]{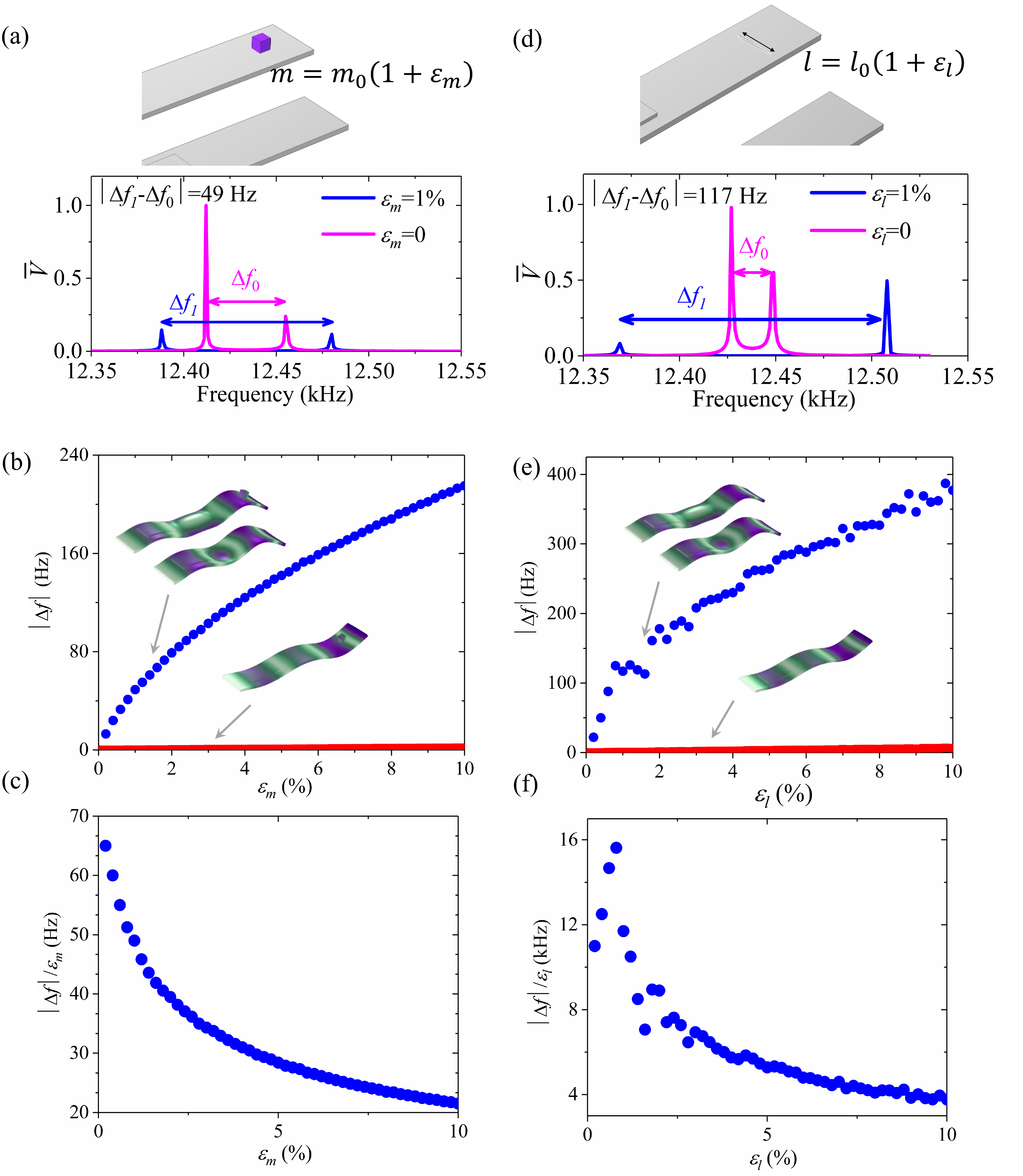}
\caption{Analysis of sensitivity of the anti-P-pseudo-Hermitian system for mechanical sensing. (a) The illustration of the system loaded by a mass and the corresponding frequency-output curves for the mass increment of $1\%$. (b) The comparison of the sensitivities of the proposed system with the conventional impedance method for mass sensing. (c) The variation of the sensitivity of the proposed system with the increase of mass perturbation. (d) The illustration of the beam with a slender surface crack and the corresponding frequency-output curve for the crack extension of $1\%$. (e) The comparison of the sensitivities of the proposed approach with the conventional impedance method for the crack growth sensing. (f) The variation of the sensitivity of the proposed approach with the increase of crack extension. }
\label{fig4}
\end{figure}

As an application example, the variation of the mass $m$ on the top surface of beam 1 is considered, as depicted in Fig. \ref{fig4}(a). The initial mass is $m_0=8$ mg and the response of the EP will be studied. It should be mentioned that the initial mass is about $0.28\%$ of the system's total mass. The fourth-order mode is selected, and the results for the first-order mode is provided in Supplementary Materials VII. First, the coupling of the system should be re-programmed to maintain its ultrasensitive performance near the EP after the loading of the mass. The corresponding transfer functions can be electrically re-tuned with values of $H=9.365$ according to Eq. (\ref{Pb}). As a result, the initial frequency splitting $\Delta f_0$ for the initial mass will be $43$Hz. Next, we characterize the bifurcation behavior of the eigenfrequencies near the EP in this system, focusing on the effects of a small mass perturbation on the initial mass. Figure \ref{fig4}(b) shows the variation of the magnitude of the incremental eigenfrequency bifurcation $\Delta f$ of the system with the normalized mass perturbation ${\epsilon_m =(m-m_0)}/{m_0}$. For comparison, the eigenfrequency shift of the single passive beam impedance method) with the mass variation is also plotted. It can be found that the eigenfrequency bifurcation will increase significantly with the increase of the mass perturbation, however, the mass perturbation is barely detected through the linear eigenfrequency change of a single passive beam. The sensitivity of the system is further quantitatively investigated through the collection of the sensing voltage in the frequency domain as illustrated in Fig. \ref{fig4}(a). The eigenfrequency splitting for the mass perturbation $1\%$ is quantitatively shown as 49 Hz. Compared with the 0.2 Hz frequency shift for the single passive beam, the enhancement factor of the sensitivity is 244. Finally, the sensitivity of the system defined as $\Delta f/\epsilon_m$ as a function of the mass perturbation is clearly interpreted in Fig. \ref{fig4}(c). It shows that the sensitivity decreases with the increase of mass perturbation. Therefore, the EP-based mass sensor is particularly ultrasensitive to tiny mass perturbation. Another factor affecting the sensitivity of the mechanical EP-based mass sensor is the physical damping of beams. When the damping coefficient is small, its influence on the system eigenfrequency is negligible. However, when the damping coefficient is large, the eigenfrequencies at different orders will not merge, which will significantly reduce the sensitivity of EP-based sensors. One advantage of our proposed system is that we can use the sensor-actuator feedback digital system to actively eliminate the negative effects induced by the physical damping, see more details in Supplementary Materials VIII.

To further demonstrate the broad capability of the proposed system, we further consider the application in the field of structural health monitoring. Our system is to monitor a tiny surface crack growth within the longer beam, as illustrated in Fig. \ref{fig4}(d). In the study, the slender crack is located 40 mm from the fixed boundary with the length being $l_0=L_1/10$, width being $L_1/160$ and depth being $h_0=L_1/250$, respectively. Following the similar procedure discussed for the mass perturbation, the transfer functions will be electrically re-tuned as 9.508 for the fourth-order modes, in particular. As a result, the initial frequency splitting $\Delta f_0$ for the initial crack is reprogrammed as 22 Hz. Figure \ref{fig4}(e) depicts the quantitative variation of the magnitude of the incremental eigenfrequency bifurcation $\Delta f$ of the system with the normalized crack growth $\epsilon_l=(l-l_0)/l_0$. The crack growth detection is significantly enhanced by comparing it with the conventional approach. Specifically, the frequency splitting $\Delta f_l$ for the crack growth $\epsilon_l=1\%$ is 139 Hz. Thus, the incremental eigenfrequency bifurcation is 117 Hz. Compared with the conventional impedance method, the sensitivity of the approach is enhanced by an order of 145 times. However, the sensitivity fluctuations occur in specific small regions due to the simultaneous reduction of mass and stiffness induced by crack growth. In these regions, both factors exhibit comparable influences on the eigenfrequency. It should be emphasized that the current approach exhibits ultra-high sensitivity to minute crack extensions, as depicted in Fig. \ref{fig4}(f), suggesting its suitability for monitoring crack growth at extremely early stages.

In summary, we have developed and analyzed an anti-P-pseudo-Hermitian mechanical system by introducing non-reciprocal coupling. This non-Hermitian system is realized in mechanical beams through the integration of piezoelectric elements connected to electrical controllers. The design incorporates a feedback control loop that enables real-time coupling operations. We theoretically demonstrate and experimentally validate the existence of tunable EPs, a key feature of non-Hermitian systems. As a practical application, we showcase the system’s effectiveness in mass sensing and minor crack growth detection. Unlike existing designs, our electrically controlled approach, governed by the transfer function, offers exceptional tunability, enabling the realization of a broad range of non-Hermitian systems, including PT-symmetric and anti-PT-symmetric configurations. This work establishes a new direction in mechanical non-Hermitian research, highlighting the role of electrically controlled coupling in structural elements and opening avenues for applications beyond mechanical sensing.

\begin{acknowledgments}
This work is supported by the Air Force Office of Scientific Research under Grant AF 9550-20-0279 with Program Manager Dr. Byung-Lip (Les) Lee.
\end{acknowledgments}


\bibliography{Reference}

\begin{thebibliography}{40}%
\makeatletter
\providecommand \@ifxundefined [1]{%
 \@ifx{#1\undefined}
}%
\providecommand \@ifnum [1]{%
 \ifnum #1\expandafter \@firstoftwo
 \else \expandafter \@secondoftwo
 \fi
}%
\providecommand \@ifx [1]{%
 \ifx #1\expandafter \@firstoftwo
 \else \expandafter \@secondoftwo
 \fi
}%
\providecommand \natexlab [1]{#1}%
\providecommand \enquote  [1]{``#1''}%
\providecommand \bibnamefont  [1]{#1}%
\providecommand \bibfnamefont [1]{#1}%
\providecommand \citenamefont [1]{#1}%
\providecommand \href@noop [0]{\@secondoftwo}%
\providecommand \href [0]{\begingroup \@sanitize@url \@href}%
\providecommand \@href[1]{\@@startlink{#1}\@@href}%
\providecommand \@@href[1]{\endgroup#1\@@endlink}%
\providecommand \@sanitize@url [0]{\catcode `\\12\catcode `\$12\catcode `\&12\catcode `\#12\catcode `\^12\catcode `\_12\catcode `\%12\relax}%
\providecommand \@@startlink[1]{}%
\providecommand \@@endlink[0]{}%
\providecommand \url  [0]{\begingroup\@sanitize@url \@url }%
\providecommand \@url [1]{\endgroup\@href {#1}{\urlprefix }}%
\providecommand \urlprefix  [0]{URL }%
\providecommand \Eprint [0]{\href }%
\providecommand \doibase [0]{https://doi.org/}%
\providecommand \selectlanguage [0]{\@gobble}%
\providecommand \bibinfo  [0]{\@secondoftwo}%
\providecommand \bibfield  [0]{\@secondoftwo}%
\providecommand \translation [1]{[#1]}%
\providecommand \BibitemOpen [0]{}%
\providecommand \bibitemStop [0]{}%
\providecommand \bibitemNoStop [0]{.\EOS\space}%
\providecommand \EOS [0]{\spacefactor3000\relax}%
\providecommand \BibitemShut  [1]{\csname bibitem#1\endcsname}%
\let\auto@bib@innerbib\@empty
\bibitem [{\citenamefont {Makris}\ \emph {et~al.}(2008)\citenamefont {Makris}, \citenamefont {El-Ganainy}, \citenamefont {Christodoulides},\ and\ \citenamefont {Musslimani}}]{makris2008beam}%
  \BibitemOpen
  \bibfield  {author} {\bibinfo {author} {\bibfnamefont {K.~G.}\ \bibnamefont {Makris}}, \bibinfo {author} {\bibfnamefont {R.}~\bibnamefont {El-Ganainy}}, \bibinfo {author} {\bibfnamefont {D.}~\bibnamefont {Christodoulides}},\ and\ \bibinfo {author} {\bibfnamefont {Z.~H.}\ \bibnamefont {Musslimani}},\ }\bibfield  {title} {\bibinfo {title} {Beam dynamics in pt symmetric optical lattices},\ }\href@noop {} {\bibfield  {journal} {\bibinfo  {journal} {Physical Review Letters}\ }\textbf {\bibinfo {volume} {100}},\ \bibinfo {pages} {103904} (\bibinfo {year} {2008})}\BibitemShut {NoStop}%
\bibitem [{\citenamefont {Mandal}\ and\ \citenamefont {Bergholtz}(2021)}]{mandal2021symmetry}%
  \BibitemOpen
  \bibfield  {author} {\bibinfo {author} {\bibfnamefont {I.}~\bibnamefont {Mandal}}\ and\ \bibinfo {author} {\bibfnamefont {E.~J.}\ \bibnamefont {Bergholtz}},\ }\bibfield  {title} {\bibinfo {title} {Symmetry and higher-order exceptional points},\ }\href@noop {} {\bibfield  {journal} {\bibinfo  {journal} {Physical review letters}\ }\textbf {\bibinfo {volume} {127}},\ \bibinfo {pages} {186601} (\bibinfo {year} {2021})}\BibitemShut {NoStop}%
\bibitem [{\citenamefont {Heiss}(2012)}]{heiss2012physics}%
  \BibitemOpen
  \bibfield  {author} {\bibinfo {author} {\bibfnamefont {W.~D.}\ \bibnamefont {Heiss}},\ }\bibfield  {title} {\bibinfo {title} {The physics of exceptional points},\ }\href@noop {} {\bibfield  {journal} {\bibinfo  {journal} {Journal of Physics A: Mathematical and Theoretical}\ }\textbf {\bibinfo {volume} {45}},\ \bibinfo {pages} {444016} (\bibinfo {year} {2012})}\BibitemShut {NoStop}%
\bibitem [{\citenamefont {Wu}\ \emph {et~al.}(2024)\citenamefont {Wu}, \citenamefont {Wang}, \citenamefont {Qian}, \citenamefont {Wang},\ and\ \citenamefont {Huang}}]{wu2024understanding}%
  \BibitemOpen
  \bibfield  {author} {\bibinfo {author} {\bibfnamefont {Q.}~\bibnamefont {Wu}}, \bibinfo {author} {\bibfnamefont {S.}~\bibnamefont {Wang}}, \bibinfo {author} {\bibfnamefont {H.}~\bibnamefont {Qian}}, \bibinfo {author} {\bibfnamefont {Y.}~\bibnamefont {Wang}},\ and\ \bibinfo {author} {\bibfnamefont {G.}~\bibnamefont {Huang}},\ }\bibfield  {title} {\bibinfo {title} {Understanding of topological mode and skin mode morphing in 1d and 2d non-hermitian resonance-based meta-lattices},\ }\href@noop {} {\bibfield  {journal} {\bibinfo  {journal} {Journal of the Mechanics and Physics of Solids}\ }\textbf {\bibinfo {volume} {193}},\ \bibinfo {pages} {105907} (\bibinfo {year} {2024})}\BibitemShut {NoStop}%
\bibitem [{\citenamefont {Bender}\ and\ \citenamefont {Boettcher}(1998)}]{bender1998real}%
  \BibitemOpen
  \bibfield  {author} {\bibinfo {author} {\bibfnamefont {C.~M.}\ \bibnamefont {Bender}}\ and\ \bibinfo {author} {\bibfnamefont {S.}~\bibnamefont {Boettcher}},\ }\bibfield  {title} {\bibinfo {title} {Real spectra in non-hermitian hamiltonians having p t symmetry},\ }\href@noop {} {\bibfield  {journal} {\bibinfo  {journal} {Physical review letters}\ }\textbf {\bibinfo {volume} {80}},\ \bibinfo {pages} {5243} (\bibinfo {year} {1998})}\BibitemShut {NoStop}%
\bibitem [{\citenamefont {Bender}\ \emph {et~al.}(2002)\citenamefont {Bender}, \citenamefont {Brody},\ and\ \citenamefont {Jones}}]{bender2002complex}%
  \BibitemOpen
  \bibfield  {author} {\bibinfo {author} {\bibfnamefont {C.~M.}\ \bibnamefont {Bender}}, \bibinfo {author} {\bibfnamefont {D.~C.}\ \bibnamefont {Brody}},\ and\ \bibinfo {author} {\bibfnamefont {H.~F.}\ \bibnamefont {Jones}},\ }\bibfield  {title} {\bibinfo {title} {Complex extension of quantum mechanics},\ }\href@noop {} {\bibfield  {journal} {\bibinfo  {journal} {Physical review letters}\ }\textbf {\bibinfo {volume} {89}},\ \bibinfo {pages} {270401} (\bibinfo {year} {2002})}\BibitemShut {NoStop}%
\bibitem [{\citenamefont {Mostafazadeh}(2003)}]{mostafazadeh2003pseudo}%
  \BibitemOpen
  \bibfield  {author} {\bibinfo {author} {\bibfnamefont {A.}~\bibnamefont {Mostafazadeh}},\ }\bibfield  {title} {\bibinfo {title} {Pseudo-hermiticity and generalized pt-and cpt-symmetries},\ }\href@noop {} {\bibfield  {journal} {\bibinfo  {journal} {Journal of Mathematical Physics}\ }\textbf {\bibinfo {volume} {44}},\ \bibinfo {pages} {974} (\bibinfo {year} {2003})}\BibitemShut {NoStop}%
\bibitem [{\citenamefont {Ashida}\ \emph {et~al.}(2020)\citenamefont {Ashida}, \citenamefont {Gong},\ and\ \citenamefont {Ueda}}]{ashida2020non}%
  \BibitemOpen
  \bibfield  {author} {\bibinfo {author} {\bibfnamefont {Y.}~\bibnamefont {Ashida}}, \bibinfo {author} {\bibfnamefont {Z.}~\bibnamefont {Gong}},\ and\ \bibinfo {author} {\bibfnamefont {M.}~\bibnamefont {Ueda}},\ }\bibfield  {title} {\bibinfo {title} {Non-hermitian physics},\ }\href@noop {} {\bibfield  {journal} {\bibinfo  {journal} {Advances in Physics}\ }\textbf {\bibinfo {volume} {69}},\ \bibinfo {pages} {249} (\bibinfo {year} {2020})}\BibitemShut {NoStop}%
\bibitem [{\citenamefont {{\"O}zdemir}\ \emph {et~al.}(2019)\citenamefont {{\"O}zdemir}, \citenamefont {Rotter}, \citenamefont {Nori},\ and\ \citenamefont {Yang}}]{ozdemir2019parity}%
  \BibitemOpen
  \bibfield  {author} {\bibinfo {author} {\bibfnamefont {{\c{S}}.~K.}\ \bibnamefont {{\"O}zdemir}}, \bibinfo {author} {\bibfnamefont {S.}~\bibnamefont {Rotter}}, \bibinfo {author} {\bibfnamefont {F.}~\bibnamefont {Nori}},\ and\ \bibinfo {author} {\bibfnamefont {L.}~\bibnamefont {Yang}},\ }\bibfield  {title} {\bibinfo {title} {Parity--time symmetry and exceptional points in photonics},\ }\href@noop {} {\bibfield  {journal} {\bibinfo  {journal} {Nature materials}\ }\textbf {\bibinfo {volume} {18}},\ \bibinfo {pages} {783} (\bibinfo {year} {2019})}\BibitemShut {NoStop}%
\bibitem [{\citenamefont {Yang}\ \emph {et~al.}(2017)\citenamefont {Yang}, \citenamefont {Liu},\ and\ \citenamefont {You}}]{yang2017anti}%
  \BibitemOpen
  \bibfield  {author} {\bibinfo {author} {\bibfnamefont {F.}~\bibnamefont {Yang}}, \bibinfo {author} {\bibfnamefont {Y.-C.}\ \bibnamefont {Liu}},\ and\ \bibinfo {author} {\bibfnamefont {L.}~\bibnamefont {You}},\ }\bibfield  {title} {\bibinfo {title} {Anti-pt symmetry in dissipatively coupled optical systems},\ }\href@noop {} {\bibfield  {journal} {\bibinfo  {journal} {Physical Review A}\ }\textbf {\bibinfo {volume} {96}},\ \bibinfo {pages} {053845} (\bibinfo {year} {2017})}\BibitemShut {NoStop}%
\bibitem [{\citenamefont {Kononchuk}\ \emph {et~al.}(2022)\citenamefont {Kononchuk}, \citenamefont {Cai}, \citenamefont {Ellis}, \citenamefont {Thevamaran},\ and\ \citenamefont {Kottos}}]{kononchuk2022exceptional}%
  \BibitemOpen
  \bibfield  {author} {\bibinfo {author} {\bibfnamefont {R.}~\bibnamefont {Kononchuk}}, \bibinfo {author} {\bibfnamefont {J.}~\bibnamefont {Cai}}, \bibinfo {author} {\bibfnamefont {F.}~\bibnamefont {Ellis}}, \bibinfo {author} {\bibfnamefont {R.}~\bibnamefont {Thevamaran}},\ and\ \bibinfo {author} {\bibfnamefont {T.}~\bibnamefont {Kottos}},\ }\bibfield  {title} {\bibinfo {title} {Exceptional-point-based accelerometers with enhanced signal-to-noise ratio},\ }\href@noop {} {\bibfield  {journal} {\bibinfo  {journal} {Nature}\ }\textbf {\bibinfo {volume} {607}},\ \bibinfo {pages} {697} (\bibinfo {year} {2022})}\BibitemShut {NoStop}%
\bibitem [{\citenamefont {Ren}\ \emph {et~al.}(2017)\citenamefont {Ren}, \citenamefont {Hodaei}, \citenamefont {Harari}, \citenamefont {Hassan}, \citenamefont {Chow}, \citenamefont {Soltani}, \citenamefont {Christodoulides},\ and\ \citenamefont {Khajavikhan}}]{ren2017ultrasensitive}%
  \BibitemOpen
  \bibfield  {author} {\bibinfo {author} {\bibfnamefont {J.}~\bibnamefont {Ren}}, \bibinfo {author} {\bibfnamefont {H.}~\bibnamefont {Hodaei}}, \bibinfo {author} {\bibfnamefont {G.}~\bibnamefont {Harari}}, \bibinfo {author} {\bibfnamefont {A.~U.}\ \bibnamefont {Hassan}}, \bibinfo {author} {\bibfnamefont {W.}~\bibnamefont {Chow}}, \bibinfo {author} {\bibfnamefont {M.}~\bibnamefont {Soltani}}, \bibinfo {author} {\bibfnamefont {D.}~\bibnamefont {Christodoulides}},\ and\ \bibinfo {author} {\bibfnamefont {M.}~\bibnamefont {Khajavikhan}},\ }\bibfield  {title} {\bibinfo {title} {Ultrasensitive micro-scale parity-time-symmetric ring laser gyroscope},\ }\href@noop {} {\bibfield  {journal} {\bibinfo  {journal} {Optics letters}\ }\textbf {\bibinfo {volume} {42}},\ \bibinfo {pages} {1556} (\bibinfo {year} {2017})}\BibitemShut {NoStop}%
\bibitem [{\citenamefont {Chen}\ \emph {et~al.}(2017)\citenamefont {Chen}, \citenamefont {Kaya~{\"O}zdemir}, \citenamefont {Zhao}, \citenamefont {Wiersig},\ and\ \citenamefont {Yang}}]{chen2017exceptional}%
  \BibitemOpen
  \bibfield  {author} {\bibinfo {author} {\bibfnamefont {W.}~\bibnamefont {Chen}}, \bibinfo {author} {\bibfnamefont {{\c{S}}.}~\bibnamefont {Kaya~{\"O}zdemir}}, \bibinfo {author} {\bibfnamefont {G.}~\bibnamefont {Zhao}}, \bibinfo {author} {\bibfnamefont {J.}~\bibnamefont {Wiersig}},\ and\ \bibinfo {author} {\bibfnamefont {L.}~\bibnamefont {Yang}},\ }\bibfield  {title} {\bibinfo {title} {Exceptional points enhance sensing in an optical microcavity},\ }\href@noop {} {\bibfield  {journal} {\bibinfo  {journal} {Nature}\ }\textbf {\bibinfo {volume} {548}},\ \bibinfo {pages} {192} (\bibinfo {year} {2017})}\BibitemShut {NoStop}%
\bibitem [{\citenamefont {De~Carlo}(2021)}]{de2021exceptional}%
  \BibitemOpen
  \bibfield  {author} {\bibinfo {author} {\bibfnamefont {M.}~\bibnamefont {De~Carlo}},\ }\bibfield  {title} {\bibinfo {title} {Exceptional points of parity-time-and anti-parity-time-symmetric devices for refractive index and absorption-based sensing},\ }\href@noop {} {\bibfield  {journal} {\bibinfo  {journal} {Results in Optics}\ }\textbf {\bibinfo {volume} {2}},\ \bibinfo {pages} {100052} (\bibinfo {year} {2021})}\BibitemShut {NoStop}%
\bibitem [{\citenamefont {Ghosh}(2021)}]{ghosh2021classical}%
  \BibitemOpen
  \bibfield  {author} {\bibinfo {author} {\bibfnamefont {P.~K.}\ \bibnamefont {Ghosh}},\ }\bibfield  {title} {\bibinfo {title} {Classical hamiltonian systems with balanced loss and gain},\ }in\ \href@noop {} {\emph {\bibinfo {booktitle} {Journal of Physics: Conference Series}}},\ Vol.\ \bibinfo {volume} {2038}\ (\bibinfo {organization} {IOP Publishing},\ \bibinfo {year} {2021})\ p.\ \bibinfo {pages} {012012}\BibitemShut {NoStop}%
\bibitem [{\citenamefont {Meng}\ \emph {et~al.}(2024)\citenamefont {Meng}, \citenamefont {Ang},\ and\ \citenamefont {Lee}}]{meng2024exceptional}%
  \BibitemOpen
  \bibfield  {author} {\bibinfo {author} {\bibfnamefont {H.}~\bibnamefont {Meng}}, \bibinfo {author} {\bibfnamefont {Y.~S.}\ \bibnamefont {Ang}},\ and\ \bibinfo {author} {\bibfnamefont {C.~H.}\ \bibnamefont {Lee}},\ }\bibfield  {title} {\bibinfo {title} {Exceptional points in non-hermitian systems: Applications and recent developments},\ }\href@noop {} {\bibfield  {journal} {\bibinfo  {journal} {Applied Physics Letters}\ }\textbf {\bibinfo {volume} {124}} (\bibinfo {year} {2024})}\BibitemShut {NoStop}%
\bibitem [{\citenamefont {R{\"u}ter}\ \emph {et~al.}(2010)\citenamefont {R{\"u}ter}, \citenamefont {Makris}, \citenamefont {El-Ganainy}, \citenamefont {Christodoulides}, \citenamefont {Segev},\ and\ \citenamefont {Kip}}]{ruter2010observation}%
  \BibitemOpen
  \bibfield  {author} {\bibinfo {author} {\bibfnamefont {C.~E.}\ \bibnamefont {R{\"u}ter}}, \bibinfo {author} {\bibfnamefont {K.~G.}\ \bibnamefont {Makris}}, \bibinfo {author} {\bibfnamefont {R.}~\bibnamefont {El-Ganainy}}, \bibinfo {author} {\bibfnamefont {D.~N.}\ \bibnamefont {Christodoulides}}, \bibinfo {author} {\bibfnamefont {M.}~\bibnamefont {Segev}},\ and\ \bibinfo {author} {\bibfnamefont {D.}~\bibnamefont {Kip}},\ }\bibfield  {title} {\bibinfo {title} {Observation of parity--time symmetry in optics},\ }\href@noop {} {\bibfield  {journal} {\bibinfo  {journal} {Nature physics}\ }\textbf {\bibinfo {volume} {6}},\ \bibinfo {pages} {192} (\bibinfo {year} {2010})}\BibitemShut {NoStop}%
\bibitem [{\citenamefont {Lin}\ \emph {et~al.}(2011)\citenamefont {Lin}, \citenamefont {Ramezani}, \citenamefont {Eichelkraut}, \citenamefont {Kottos}, \citenamefont {Cao},\ and\ \citenamefont {Christodoulides}}]{lin2011unidirectional}%
  \BibitemOpen
  \bibfield  {author} {\bibinfo {author} {\bibfnamefont {Z.}~\bibnamefont {Lin}}, \bibinfo {author} {\bibfnamefont {H.}~\bibnamefont {Ramezani}}, \bibinfo {author} {\bibfnamefont {T.}~\bibnamefont {Eichelkraut}}, \bibinfo {author} {\bibfnamefont {T.}~\bibnamefont {Kottos}}, \bibinfo {author} {\bibfnamefont {H.}~\bibnamefont {Cao}},\ and\ \bibinfo {author} {\bibfnamefont {D.~N.}\ \bibnamefont {Christodoulides}},\ }\bibfield  {title} {\bibinfo {title} {Unidirectional invisibility induced by pt-symmetric periodic structures},\ }\href@noop {} {\bibfield  {journal} {\bibinfo  {journal} {Physical Review Letters}\ }\textbf {\bibinfo {volume} {106}},\ \bibinfo {pages} {213901} (\bibinfo {year} {2011})}\BibitemShut {NoStop}%
\bibitem [{\citenamefont {De~Carlo}\ \emph {et~al.}(2022)\citenamefont {De~Carlo}, \citenamefont {De~Leonardis}, \citenamefont {Soref}, \citenamefont {Colatorti},\ and\ \citenamefont {Passaro}}]{de2022non}%
  \BibitemOpen
  \bibfield  {author} {\bibinfo {author} {\bibfnamefont {M.}~\bibnamefont {De~Carlo}}, \bibinfo {author} {\bibfnamefont {F.}~\bibnamefont {De~Leonardis}}, \bibinfo {author} {\bibfnamefont {R.~A.}\ \bibnamefont {Soref}}, \bibinfo {author} {\bibfnamefont {L.}~\bibnamefont {Colatorti}},\ and\ \bibinfo {author} {\bibfnamefont {V.~M.}\ \bibnamefont {Passaro}},\ }\bibfield  {title} {\bibinfo {title} {Non-hermitian sensing in photonics and electronics: A review},\ }\href@noop {} {\bibfield  {journal} {\bibinfo  {journal} {Sensors}\ }\textbf {\bibinfo {volume} {22}},\ \bibinfo {pages} {3977} (\bibinfo {year} {2022})}\BibitemShut {NoStop}%
\bibitem [{\citenamefont {Hodaei}\ \emph {et~al.}(2017)\citenamefont {Hodaei}, \citenamefont {Hassan}, \citenamefont {Wittek}, \citenamefont {Garcia-Gracia}, \citenamefont {El-Ganainy}, \citenamefont {Christodoulides},\ and\ \citenamefont {Khajavikhan}}]{hodaei2017enhanced}%
  \BibitemOpen
  \bibfield  {author} {\bibinfo {author} {\bibfnamefont {H.}~\bibnamefont {Hodaei}}, \bibinfo {author} {\bibfnamefont {A.~U.}\ \bibnamefont {Hassan}}, \bibinfo {author} {\bibfnamefont {S.}~\bibnamefont {Wittek}}, \bibinfo {author} {\bibfnamefont {H.}~\bibnamefont {Garcia-Gracia}}, \bibinfo {author} {\bibfnamefont {R.}~\bibnamefont {El-Ganainy}}, \bibinfo {author} {\bibfnamefont {D.~N.}\ \bibnamefont {Christodoulides}},\ and\ \bibinfo {author} {\bibfnamefont {M.}~\bibnamefont {Khajavikhan}},\ }\bibfield  {title} {\bibinfo {title} {Enhanced sensitivity at higher-order exceptional points},\ }\href@noop {} {\bibfield  {journal} {\bibinfo  {journal} {Nature}\ }\textbf {\bibinfo {volume} {548}},\ \bibinfo {pages} {187} (\bibinfo {year} {2017})}\BibitemShut {NoStop}%
\bibitem [{\citenamefont {Rosa}\ \emph {et~al.}(2021)\citenamefont {Rosa}, \citenamefont {Mazzotti},\ and\ \citenamefont {Ruzzene}}]{rosa2021exceptional}%
  \BibitemOpen
  \bibfield  {author} {\bibinfo {author} {\bibfnamefont {M.~I.}\ \bibnamefont {Rosa}}, \bibinfo {author} {\bibfnamefont {M.}~\bibnamefont {Mazzotti}},\ and\ \bibinfo {author} {\bibfnamefont {M.}~\bibnamefont {Ruzzene}},\ }\bibfield  {title} {\bibinfo {title} {Exceptional points and enhanced sensitivity in pt-symmetric continuous elastic media},\ }\href@noop {} {\bibfield  {journal} {\bibinfo  {journal} {Journal of the Mechanics and Physics of Solids}\ }\textbf {\bibinfo {volume} {149}},\ \bibinfo {pages} {104325} (\bibinfo {year} {2021})}\BibitemShut {NoStop}%
\bibitem [{\citenamefont {Wu}\ \emph {et~al.}(2019)\citenamefont {Wu}, \citenamefont {Chen},\ and\ \citenamefont {Huang}}]{wu2019asymmetric}%
  \BibitemOpen
  \bibfield  {author} {\bibinfo {author} {\bibfnamefont {Q.}~\bibnamefont {Wu}}, \bibinfo {author} {\bibfnamefont {Y.}~\bibnamefont {Chen}},\ and\ \bibinfo {author} {\bibfnamefont {G.}~\bibnamefont {Huang}},\ }\bibfield  {title} {\bibinfo {title} {Asymmetric scattering of flexural waves in a parity-time symmetric metamaterial beam},\ }\href@noop {} {\bibfield  {journal} {\bibinfo  {journal} {The Journal of the Acoustical Society of America}\ }\textbf {\bibinfo {volume} {146}},\ \bibinfo {pages} {850} (\bibinfo {year} {2019})}\BibitemShut {NoStop}%
\bibitem [{\citenamefont {Thomes}\ \emph {et~al.}(2024)\citenamefont {Thomes}, \citenamefont {Rosa},\ and\ \citenamefont {Erturk}}]{thomes2024experimental}%
  \BibitemOpen
  \bibfield  {author} {\bibinfo {author} {\bibfnamefont {R.~L.}\ \bibnamefont {Thomes}}, \bibinfo {author} {\bibfnamefont {M.~I.}\ \bibnamefont {Rosa}},\ and\ \bibinfo {author} {\bibfnamefont {A.}~\bibnamefont {Erturk}},\ }\bibfield  {title} {\bibinfo {title} {Experimental realization of tunable exceptional points in a resonant non-hermitian piezoelectrically coupled waveguide},\ }\href@noop {} {\bibfield  {journal} {\bibinfo  {journal} {Applied Physics Letters}\ }\textbf {\bibinfo {volume} {124}} (\bibinfo {year} {2024})}\BibitemShut {NoStop}%
\bibitem [{\citenamefont {Xu}\ \emph {et~al.}(2015)\citenamefont {Xu}, \citenamefont {Liu}, \citenamefont {Sun},\ and\ \citenamefont {Li}}]{xu2015mechanical}%
  \BibitemOpen
  \bibfield  {author} {\bibinfo {author} {\bibfnamefont {X.-W.}\ \bibnamefont {Xu}}, \bibinfo {author} {\bibfnamefont {Y.-x.}\ \bibnamefont {Liu}}, \bibinfo {author} {\bibfnamefont {C.-P.}\ \bibnamefont {Sun}},\ and\ \bibinfo {author} {\bibfnamefont {Y.}~\bibnamefont {Li}},\ }\bibfield  {title} {\bibinfo {title} {Mechanical pt symmetry in coupled optomechanical systems},\ }\href@noop {} {\bibfield  {journal} {\bibinfo  {journal} {Physical Review A}\ }\textbf {\bibinfo {volume} {92}},\ \bibinfo {pages} {013852} (\bibinfo {year} {2015})}\BibitemShut {NoStop}%
\bibitem [{\citenamefont {Jin}\ \emph {et~al.}(2024)\citenamefont {Jin}, \citenamefont {Li}, \citenamefont {Djafari-Rouhani}, \citenamefont {Li},\ and\ \citenamefont {Xiang}}]{jin2024exceptional}%
  \BibitemOpen
  \bibfield  {author} {\bibinfo {author} {\bibfnamefont {Y.}~\bibnamefont {Jin}}, \bibinfo {author} {\bibfnamefont {W.}~\bibnamefont {Li}}, \bibinfo {author} {\bibfnamefont {B.}~\bibnamefont {Djafari-Rouhani}}, \bibinfo {author} {\bibfnamefont {Y.}~\bibnamefont {Li}},\ and\ \bibinfo {author} {\bibfnamefont {Y.}~\bibnamefont {Xiang}},\ }\bibfield  {title} {\bibinfo {title} {Exceptional points for crack detection in non-hermitian beams},\ }\href@noop {} {\bibfield  {journal} {\bibinfo  {journal} {Journal of Sound and Vibration}\ }\textbf {\bibinfo {volume} {572}},\ \bibinfo {pages} {118162} (\bibinfo {year} {2024})}\BibitemShut {NoStop}%
\bibitem [{\citenamefont {Cai}\ \emph {et~al.}(2022)\citenamefont {Cai}, \citenamefont {Jin}, \citenamefont {Li}, \citenamefont {Rabczuk}, \citenamefont {Pennec}, \citenamefont {Djafari-Rouhani},\ and\ \citenamefont {Zhuang}}]{cai2022exceptional}%
  \BibitemOpen
  \bibfield  {author} {\bibinfo {author} {\bibfnamefont {R.}~\bibnamefont {Cai}}, \bibinfo {author} {\bibfnamefont {Y.}~\bibnamefont {Jin}}, \bibinfo {author} {\bibfnamefont {Y.}~\bibnamefont {Li}}, \bibinfo {author} {\bibfnamefont {T.}~\bibnamefont {Rabczuk}}, \bibinfo {author} {\bibfnamefont {Y.}~\bibnamefont {Pennec}}, \bibinfo {author} {\bibfnamefont {B.}~\bibnamefont {Djafari-Rouhani}},\ and\ \bibinfo {author} {\bibfnamefont {X.}~\bibnamefont {Zhuang}},\ }\bibfield  {title} {\bibinfo {title} {Exceptional points and skin modes in non-hermitian metabeams},\ }\href@noop {} {\bibfield  {journal} {\bibinfo  {journal} {Physical Review Applied}\ }\textbf {\bibinfo {volume} {18}},\ \bibinfo {pages} {014067} (\bibinfo {year} {2022})}\BibitemShut {NoStop}%
\bibitem [{\citenamefont {Yuan}\ \emph {et~al.}(2022)\citenamefont {Yuan}, \citenamefont {Geng}, \citenamefont {Huang}, \citenamefont {Guo}, \citenamefont {Yang}, \citenamefont {Hu},\ and\ \citenamefont {Zhou}}]{yuan2022exceptional}%
  \BibitemOpen
  \bibfield  {author} {\bibinfo {author} {\bibfnamefont {J.}~\bibnamefont {Yuan}}, \bibinfo {author} {\bibfnamefont {L.}~\bibnamefont {Geng}}, \bibinfo {author} {\bibfnamefont {J.}~\bibnamefont {Huang}}, \bibinfo {author} {\bibfnamefont {Q.}~\bibnamefont {Guo}}, \bibinfo {author} {\bibfnamefont {J.}~\bibnamefont {Yang}}, \bibinfo {author} {\bibfnamefont {G.}~\bibnamefont {Hu}},\ and\ \bibinfo {author} {\bibfnamefont {X.}~\bibnamefont {Zhou}},\ }\bibfield  {title} {\bibinfo {title} {Exceptional points induced by time-varying mass to enhance the sensitivity of defect detection},\ }\href@noop {} {\bibfield  {journal} {\bibinfo  {journal} {Physical Review Applied}\ }\textbf {\bibinfo {volume} {18}},\ \bibinfo {pages} {064055} (\bibinfo {year} {2022})}\BibitemShut {NoStop}%
\bibitem [{\citenamefont {Fan}\ \emph {et~al.}(2020)\citenamefont {Fan}, \citenamefont {Chen}, \citenamefont {Zhao}, \citenamefont {Wen},\ and\ \citenamefont {Huang}}]{fan2020antiparity}%
  \BibitemOpen
  \bibfield  {author} {\bibinfo {author} {\bibfnamefont {H.}~\bibnamefont {Fan}}, \bibinfo {author} {\bibfnamefont {J.}~\bibnamefont {Chen}}, \bibinfo {author} {\bibfnamefont {Z.}~\bibnamefont {Zhao}}, \bibinfo {author} {\bibfnamefont {J.}~\bibnamefont {Wen}},\ and\ \bibinfo {author} {\bibfnamefont {Y.-P.}\ \bibnamefont {Huang}},\ }\bibfield  {title} {\bibinfo {title} {Antiparity-time symmetry in passive nanophotonics},\ }\href@noop {} {\bibfield  {journal} {\bibinfo  {journal} {ACS photonics}\ }\textbf {\bibinfo {volume} {7}},\ \bibinfo {pages} {3035} (\bibinfo {year} {2020})}\BibitemShut {NoStop}%
\bibitem [{\citenamefont {De~Carlo}\ \emph {et~al.}(2019)\citenamefont {De~Carlo}, \citenamefont {De~Leonardis}, \citenamefont {Lamberti},\ and\ \citenamefont {Passaro}}]{de2019high}%
  \BibitemOpen
  \bibfield  {author} {\bibinfo {author} {\bibfnamefont {M.}~\bibnamefont {De~Carlo}}, \bibinfo {author} {\bibfnamefont {F.}~\bibnamefont {De~Leonardis}}, \bibinfo {author} {\bibfnamefont {L.}~\bibnamefont {Lamberti}},\ and\ \bibinfo {author} {\bibfnamefont {V.~M.}\ \bibnamefont {Passaro}},\ }\bibfield  {title} {\bibinfo {title} {High-sensitivity real-splitting anti-pt-symmetric microscale optical gyroscope},\ }\href@noop {} {\bibfield  {journal} {\bibinfo  {journal} {Optics letters}\ }\textbf {\bibinfo {volume} {44}},\ \bibinfo {pages} {3956} (\bibinfo {year} {2019})}\BibitemShut {NoStop}%
\bibitem [{\citenamefont {Li}\ \emph {et~al.}(2019)\citenamefont {Li}, \citenamefont {Peng}, \citenamefont {Han}, \citenamefont {Miri}, \citenamefont {Li}, \citenamefont {Xiao}, \citenamefont {Zhu}, \citenamefont {Zhao}, \citenamefont {Al{\`u}}, \citenamefont {Fan} \emph {et~al.}}]{li2019anti}%
  \BibitemOpen
  \bibfield  {author} {\bibinfo {author} {\bibfnamefont {Y.}~\bibnamefont {Li}}, \bibinfo {author} {\bibfnamefont {Y.-G.}\ \bibnamefont {Peng}}, \bibinfo {author} {\bibfnamefont {L.}~\bibnamefont {Han}}, \bibinfo {author} {\bibfnamefont {M.-A.}\ \bibnamefont {Miri}}, \bibinfo {author} {\bibfnamefont {W.}~\bibnamefont {Li}}, \bibinfo {author} {\bibfnamefont {M.}~\bibnamefont {Xiao}}, \bibinfo {author} {\bibfnamefont {X.-F.}\ \bibnamefont {Zhu}}, \bibinfo {author} {\bibfnamefont {J.}~\bibnamefont {Zhao}}, \bibinfo {author} {\bibfnamefont {A.}~\bibnamefont {Al{\`u}}}, \bibinfo {author} {\bibfnamefont {S.}~\bibnamefont {Fan}}, \emph {et~al.},\ }\bibfield  {title} {\bibinfo {title} {Anti--parity-time symmetry in diffusive systems},\ }\href@noop {} {\bibfield  {journal} {\bibinfo  {journal} {Science}\ }\textbf {\bibinfo {volume} {364}},\ \bibinfo {pages} {170} (\bibinfo {year} {2019})}\BibitemShut {NoStop}%
\bibitem [{\citenamefont {Choi}\ \emph {et~al.}(2018)\citenamefont {Choi}, \citenamefont {Hahn}, \citenamefont {Yoon},\ and\ \citenamefont {Song}}]{choi2018observation}%
  \BibitemOpen
  \bibfield  {author} {\bibinfo {author} {\bibfnamefont {Y.}~\bibnamefont {Choi}}, \bibinfo {author} {\bibfnamefont {C.}~\bibnamefont {Hahn}}, \bibinfo {author} {\bibfnamefont {J.~W.}\ \bibnamefont {Yoon}},\ and\ \bibinfo {author} {\bibfnamefont {S.~H.}\ \bibnamefont {Song}},\ }\bibfield  {title} {\bibinfo {title} {Observation of an anti-pt-symmetric exceptional point and energy-difference conserving dynamics in electrical circuit resonators},\ }\href@noop {} {\bibfield  {journal} {\bibinfo  {journal} {Nature communications}\ }\textbf {\bibinfo {volume} {9}},\ \bibinfo {pages} {2182} (\bibinfo {year} {2018})}\BibitemShut {NoStop}%
\bibitem [{\citenamefont {Zhang}\ \emph {et~al.}(2020)\citenamefont {Zhang}, \citenamefont {Feng}, \citenamefont {Chen}, \citenamefont {Ge},\ and\ \citenamefont {Wan}}]{zhang2020synthetic}%
  \BibitemOpen
  \bibfield  {author} {\bibinfo {author} {\bibfnamefont {F.}~\bibnamefont {Zhang}}, \bibinfo {author} {\bibfnamefont {Y.}~\bibnamefont {Feng}}, \bibinfo {author} {\bibfnamefont {X.}~\bibnamefont {Chen}}, \bibinfo {author} {\bibfnamefont {L.}~\bibnamefont {Ge}},\ and\ \bibinfo {author} {\bibfnamefont {W.}~\bibnamefont {Wan}},\ }\bibfield  {title} {\bibinfo {title} {Synthetic anti-pt symmetry in a single microcavity},\ }\href@noop {} {\bibfield  {journal} {\bibinfo  {journal} {Physical review letters}\ }\textbf {\bibinfo {volume} {124}},\ \bibinfo {pages} {053901} (\bibinfo {year} {2020})}\BibitemShut {NoStop}%
\bibitem [{\citenamefont {Wei}\ \emph {et~al.}(2021)\citenamefont {Wei}, \citenamefont {Zhou}, \citenamefont {Chen}, \citenamefont {Ding}, \citenamefont {Dong},\ and\ \citenamefont {Zhang}}]{wei2021anti}%
  \BibitemOpen
  \bibfield  {author} {\bibinfo {author} {\bibfnamefont {Y.}~\bibnamefont {Wei}}, \bibinfo {author} {\bibfnamefont {H.}~\bibnamefont {Zhou}}, \bibinfo {author} {\bibfnamefont {Y.}~\bibnamefont {Chen}}, \bibinfo {author} {\bibfnamefont {Y.}~\bibnamefont {Ding}}, \bibinfo {author} {\bibfnamefont {J.}~\bibnamefont {Dong}},\ and\ \bibinfo {author} {\bibfnamefont {X.}~\bibnamefont {Zhang}},\ }\bibfield  {title} {\bibinfo {title} {Anti-parity-time symmetry enabled on-chip chiral polarizer},\ }\href@noop {} {\bibfield  {journal} {\bibinfo  {journal} {Photonics Research}\ }\textbf {\bibinfo {volume} {10}},\ \bibinfo {pages} {76} (\bibinfo {year} {2021})}\BibitemShut {NoStop}%
\bibitem [{\citenamefont {Zhang}\ \emph {et~al.}(2019)\citenamefont {Zhang}, \citenamefont {Jiang},\ and\ \citenamefont {Chan}}]{zhang2019dynamically}%
  \BibitemOpen
  \bibfield  {author} {\bibinfo {author} {\bibfnamefont {X.-L.}\ \bibnamefont {Zhang}}, \bibinfo {author} {\bibfnamefont {T.}~\bibnamefont {Jiang}},\ and\ \bibinfo {author} {\bibfnamefont {C.~T.}\ \bibnamefont {Chan}},\ }\bibfield  {title} {\bibinfo {title} {Dynamically encircling an exceptional point in anti-parity-time symmetric systems: asymmetric mode switching for symmetry-broken modes},\ }\href@noop {} {\bibfield  {journal} {\bibinfo  {journal} {Light: Science \& Applications}\ }\textbf {\bibinfo {volume} {8}},\ \bibinfo {pages} {88} (\bibinfo {year} {2019})}\BibitemShut {NoStop}%
\bibitem [{\citenamefont {Zheng}\ \emph {et~al.}(2020)\citenamefont {Zheng}, \citenamefont {Tian}, \citenamefont {Li}, \citenamefont {Wen}, \citenamefont {Wei},\ and\ \citenamefont {Li}}]{zheng2020efficient}%
  \BibitemOpen
  \bibfield  {author} {\bibinfo {author} {\bibfnamefont {C.}~\bibnamefont {Zheng}}, \bibinfo {author} {\bibfnamefont {J.}~\bibnamefont {Tian}}, \bibinfo {author} {\bibfnamefont {D.}~\bibnamefont {Li}}, \bibinfo {author} {\bibfnamefont {J.}~\bibnamefont {Wen}}, \bibinfo {author} {\bibfnamefont {S.}~\bibnamefont {Wei}},\ and\ \bibinfo {author} {\bibfnamefont {Y.}~\bibnamefont {Li}},\ }\bibfield  {title} {\bibinfo {title} {Efficient quantum simulation of an anti-p-pseudo-hermitian two-level system},\ }\href@noop {} {\bibfield  {journal} {\bibinfo  {journal} {Entropy}\ }\textbf {\bibinfo {volume} {22}},\ \bibinfo {pages} {812} (\bibinfo {year} {2020})}\BibitemShut {NoStop}%
\bibitem [{\citenamefont {Chen}\ \emph {et~al.}(2021)\citenamefont {Chen}, \citenamefont {Li}, \citenamefont {Scheibner}, \citenamefont {Vitelli},\ and\ \citenamefont {Huang}}]{chen2021realization}%
  \BibitemOpen
  \bibfield  {author} {\bibinfo {author} {\bibfnamefont {Y.}~\bibnamefont {Chen}}, \bibinfo {author} {\bibfnamefont {X.}~\bibnamefont {Li}}, \bibinfo {author} {\bibfnamefont {C.}~\bibnamefont {Scheibner}}, \bibinfo {author} {\bibfnamefont {V.}~\bibnamefont {Vitelli}},\ and\ \bibinfo {author} {\bibfnamefont {G.}~\bibnamefont {Huang}},\ }\bibfield  {title} {\bibinfo {title} {Realization of active metamaterials with odd micropolar elasticity},\ }\href@noop {} {\bibfield  {journal} {\bibinfo  {journal} {Nature communications}\ }\textbf {\bibinfo {volume} {12}},\ \bibinfo {pages} {5935} (\bibinfo {year} {2021})}\BibitemShut {NoStop}%
\bibitem [{\citenamefont {Wang}\ \emph {et~al.}(2024)\citenamefont {Wang}, \citenamefont {Wu}, \citenamefont {Tian},\ and\ \citenamefont {Huang}}]{wang2024non}%
  \BibitemOpen
  \bibfield  {author} {\bibinfo {author} {\bibfnamefont {Y.}~\bibnamefont {Wang}}, \bibinfo {author} {\bibfnamefont {Q.}~\bibnamefont {Wu}}, \bibinfo {author} {\bibfnamefont {Y.}~\bibnamefont {Tian}},\ and\ \bibinfo {author} {\bibfnamefont {G.}~\bibnamefont {Huang}},\ }\bibfield  {title} {\bibinfo {title} {Non-hermitian wave dynamics of odd plates: Microstructure design and theoretical modelling},\ }\href@noop {} {\bibfield  {journal} {\bibinfo  {journal} {Journal of the Mechanics and Physics of Solids}\ }\textbf {\bibinfo {volume} {182}},\ \bibinfo {pages} {105462} (\bibinfo {year} {2024})}\BibitemShut {NoStop}%
\bibitem [{\citenamefont {Wu}\ \emph {et~al.}(2023{\natexlab{a}})\citenamefont {Wu}, \citenamefont {Shivashankar}, \citenamefont {Xu}, \citenamefont {Chen},\ and\ \citenamefont {Huang}}]{wu2023engineering}%
  \BibitemOpen
  \bibfield  {author} {\bibinfo {author} {\bibfnamefont {Q.}~\bibnamefont {Wu}}, \bibinfo {author} {\bibfnamefont {P.}~\bibnamefont {Shivashankar}}, \bibinfo {author} {\bibfnamefont {X.}~\bibnamefont {Xu}}, \bibinfo {author} {\bibfnamefont {Y.}~\bibnamefont {Chen}},\ and\ \bibinfo {author} {\bibfnamefont {G.}~\bibnamefont {Huang}},\ }\bibfield  {title} {\bibinfo {title} {Engineering nonreciprocal wave dispersion in a nonlocal micropolar metabeam},\ }\href@noop {} {\bibfield  {journal} {\bibinfo  {journal} {Journal of Composite Materials}\ }\textbf {\bibinfo {volume} {57}},\ \bibinfo {pages} {771} (\bibinfo {year} {2023}{\natexlab{a}})}\BibitemShut {NoStop}%
\bibitem [{\citenamefont {Wu}\ \emph {et~al.}(2023{\natexlab{b}})\citenamefont {Wu}, \citenamefont {Xu}, \citenamefont {Qian}, \citenamefont {Wang}, \citenamefont {Zhu}, \citenamefont {Yan}, \citenamefont {Ma}, \citenamefont {Chen},\ and\ \citenamefont {Huang}}]{wu2023active}%
  \BibitemOpen
  \bibfield  {author} {\bibinfo {author} {\bibfnamefont {Q.}~\bibnamefont {Wu}}, \bibinfo {author} {\bibfnamefont {X.}~\bibnamefont {Xu}}, \bibinfo {author} {\bibfnamefont {H.}~\bibnamefont {Qian}}, \bibinfo {author} {\bibfnamefont {S.}~\bibnamefont {Wang}}, \bibinfo {author} {\bibfnamefont {R.}~\bibnamefont {Zhu}}, \bibinfo {author} {\bibfnamefont {Z.}~\bibnamefont {Yan}}, \bibinfo {author} {\bibfnamefont {H.}~\bibnamefont {Ma}}, \bibinfo {author} {\bibfnamefont {Y.}~\bibnamefont {Chen}},\ and\ \bibinfo {author} {\bibfnamefont {G.}~\bibnamefont {Huang}},\ }\bibfield  {title} {\bibinfo {title} {Active metamaterials for realizing odd mass density},\ }\href@noop {} {\bibfield  {journal} {\bibinfo  {journal} {Proceedings of the National Academy of Sciences}\ }\textbf {\bibinfo {volume} {120}},\ \bibinfo {pages} {e2209829120} (\bibinfo {year} {2023}{\natexlab{b}})}\BibitemShut {NoStop}%
\bibitem [{\citenamefont {Strang}(2022)}]{strang2022introduction}%
  \BibitemOpen
  \bibfield  {author} {\bibinfo {author} {\bibfnamefont {G.}~\bibnamefont {Strang}},\ }\href@noop {} {\emph {\bibinfo {title} {Introduction to linear algebra}}}\ (\bibinfo  {publisher} {SIAM},\ \bibinfo {year} {2022})\BibitemShut {NoStop}%
\end{thebibliography}%

\end{document}